\documentclass[12pt]{article}
\usepackage{amsfonts,amssymb,epsfig,amsmath,mathtools}
\usepackage{color}

\usepackage[enableskew]{youngtab}
\usepackage{mathrsfs}
\usepackage{graphicx}
\usepackage{subcaption}

\renewcommand{\baselinestretch}{1.2}

\newcommand{\be}{\begin{eqnarray}}
\newcommand{\ee}{\end{eqnarray}}

\newcommand{\bn}{\begin{enumerate}}
\newcommand{\en}{\end{enumerate}}

\begin{document}

\makeatletter \@addtoreset{equation}{section} \makeatother
\renewcommand{\theequation}{\thesection.\arabic{equation}}
\renewcommand{\thefootnote}{\alph{footnote}}

\begin{titlepage}

\begin{center}
\hfill {\tt SNUTP18-007}\\
\hfill{\tt KIAS-P18106}\\

\vspace{2cm}

{\Large\bf Comments on deconfinement in AdS/CFT}

\vspace{2cm}

\renewcommand{\thefootnote}{\alph{footnote}}

{\large Sunjin Choi$^1$, Joonho Kim$^2$, Seok Kim$^1$ and June Nahmgoong$^1$}

\vspace{0.7cm}

\textit{$^1$Department of Physics and Astronomy \& Center for
Theoretical Physics,\\
Seoul National University, Seoul 08826, Korea.}\\

\vspace{0.2cm}

\textit{$^2$School of Physics, Korea Institute for Advanced Study,
Seoul 02455, Korea.}\\

\vspace{0.7cm}

E-mails: {\tt csj37100@snu.ac.kr, joonhokim@kias.re.kr, \\
skim@phya.snu.ac.kr, earendil25@snu.ac.kr}

\end{center}

\vspace{1cm}

\begin{abstract}

We study the index of $\mathcal{N}=4$ Yang-Mills theory on
$S^3\times\mathbb{R}$. We argue that the index should undergo a large $N$ 
deconfinement phase transition, by computing an upper bound of its `temperature.'
We compute this bound by optimizing the phases of fugacities.
The bound we find has some features analogous to the Hagedorn temperature. We briefly
discuss a possible mechanism of the actual deconfinement transition below our bound.
Our upper bound is lower than the Hawking-Page
transition `temperature' of known BPS black holes in the AdS$_5$ dual.
We thus expect the existence of new black holes.

\end{abstract}

\end{titlepage}

\renewcommand{\thefootnote}{\arabic{footnote}}

\setcounter{footnote}{0}

\renewcommand{\baselinestretch}{1}

\tableofcontents

\renewcommand{\baselinestretch}{1.2}

\section{Introduction}

Anti de Sitter (AdS) spacetime is an ideal setting to study various fundamental issues of
quantum gravity. Among others, we shall use AdS/CFT \cite{Maldacena:1997re}
to study aspects of black hole thermodynamics microscopically. An important feature of black hole
thermodynamics in AdS is the Hawking-Page transition \cite{Hawking:1982dh}, between large
black holes and thermal gravitons. In the CFT dual, this was suggested to be the
confinement-deconfinement transition \cite{Witten:1998zw} on a sphere.
Its details have been studied in weakly coupled gauge theory \cite{Aharony:2003sx}.
Even at weak-coupling, many qualitative features are
similar to what one expects at strong coupling from black holes.

One hopes to make these studies more quantitative in supersymmetric models,
in a more tractable sector preserving some SUSY. We know
a class of supersymmetric black hole solutions in $AdS_5\times S^5$
\cite{Gutowski:2004ez,Gutowski:2004yv,Chong:2005da,Kunduri:2006ek}.
Furthermore, numerical evidences are being
found that there exist more general `hairy' BPS black holes
\cite{Markeviciute:2018yal,Markeviciute:2018cqs}. In the
canonical ensemble (or grand canonical ensemble in BPS sector), we expect that
all these black hole saddle points are thermodynamically sub-dominant
than thermal gravitons at low temperature.\footnote{The
Bekenstein-Hawking temperatures of BPS black holes are zero. In this paper, by
`temperature' we mean inverse chemical potentials conjugate to charges which are responsible for
the BPS energy.} Increasing the temperature, one of these black holes will
start to dominate over thermal gravitons after its Hawking-Page transition.
This black hole will set the deconfinement `temperature' in the BPS sector of the CFT dual.

BPS sectors at strong coupling are easily studied using Witten indices, but
at the risk of possible boson/fermion cancelations. Such indices for SCFTs on
$S^3\times\mathbb{R}$ were discussed in
\cite{Kinney:2005ej,Romelsberger:2005eg}. However, after the studies of \cite{Kinney:2005ej},
it has been believed that the black hole physics is invisible in the index.
An apparent technical reason seemed to be severe
boson/fermion cancelations. Let us discuss in more detail what this possibly means.
Consider the inverse Laplace transformation of an index 
$\sum_j\Omega_j x^j$,
at a macroscopic charge $j\sim N^2$ in the large $N$ limit. This is a schematic
expression: as presented in section 2, the index can have
more charges and fugacities compatible with certain supercharges. Still, the charges
of our interest are all angular momenta in $AdS_5\times S^5$, whose linear combinations
will play the role of $j$ above. $\Omega_j$'s are positive or negative integers, counted
with $-1$ sign for fermions. The alternations of $\pm$ signs can be quite
random. For instance, the index of the $\mathcal{N}=4$ Yang-Mills with $U(2)$ gauge group
is given by
\begin{eqnarray}
  \hspace*{-1cm}&&
  1 + 3 x^2 - 2 x^3 + 9 x^4 - 6 x^5 + 11 x^6 - 6 x^7 + 9 x^8 + 14 x^9 -
  21 x^{10} + 36 x^{11} - 17 x^{12} - 18 x^{13}\nonumber\\
  \hspace*{-1cm}&&
   + 114 x^{14} - 194 x^{15} + 258 x^{16} - 168 x^{17} - 112 x^{18} + 630 x^{19}
  - 1089 x^{20} + 1130 x^{21} - 273 x^{22}\nonumber\\
  \hspace*{-1cm}&&
  - 1632 x^{23} + 4104 x^{24} - 5364 x^{25} + 3426 x^{26} +
  3152 x^{27} - 13233 x^{28} + 21336 x^{29} - 18319 x^{30}\nonumber\\
  \hspace*{-1cm}&&
  - 2994 x^{31} + 40752 x^{32} - 76884 x^{33} + 78012 x^{34} - 11808 x^{35}
  +\cdots\ .
\end{eqnarray}
See section 2, above (\ref{index-large-N}), for our definition of $x$ and $j$ here.
Although this result is not relevant for either large $N$ or large charge
macroscopic approximation, it illustrates random alternations of signs
as $j$ increases by its quantized unit.

Here we would like to comment that there could be two stages
at which boson/fermion cancelation can happen. First is the intrinsic cancelation
within a given $\Omega_j$, due to $(-1)^F$. If
$\log\Omega_j\sim\mathcal{O}(N^0)$ even at $j\sim N^2$, then the index would not
be useful for studying black holes. However, suppose
the case in which each $\Omega_j$ is macroscopic. Even in this case,
a macroscopic saddle point approximation of the inverse Laplace transformation could see
apparently much smaller degeneracy than each $|\Omega_j|$.
Naively performing the macroscopic saddle point approximation,
the quantized nature of the charges will be highly obscured. E.g. one cannot
precisely say whether one is counting the level $j$, or $j\pm 1$, or $j\pm 2$, and so on.
So there is a potential chance that each term in the series
$\sum_j\Omega_j x^j$ exhibits macroscopic entropy, while
a saddle point approximation captures certain nearby terms smeared out, thus
looking trivial.

Recently, a possibility of improving the latter situation was found in
\cite{Choi:2018hmj}. The simple idea is to turn on the imaginary parts of
chemical potentials, and tune them to optimally obstruct boson/fermion
cancelations (or smearing) at nearby macroscopic charges. This possibility was first
noticed in \cite{Choi:2018hmj} by inspecting the extremized chemical potentials in an
`entropy function' \cite{Hosseini:2017mds} for known BPS
black hole solutions, realizing that they have substantial imaginary parts.
This will yield extra phase factors in the fugacity expansion, which hopefully may
be tuned to tame the rapid $\pm$ alternations of nearby terms. For instance, introduction
of such phases in the index to obstruct nearby states' cancelation appeared
in \cite{Kim:2011mv}. With this idea, the large black hole limit of
\cite{Gutowski:2004ez,Gutowski:2004yv,Chong:2005da,Kunduri:2006ek} was successfully
studied microscopically by analyzing a Cardy-like limit of the index, including the
counting of their microstates \cite{Choi:2018hmj}. Away from the large black hole limit,
the possibility of more nontrivial black holes than the known analytic solutions was also
discussed \cite{Choi:2018hmj}. So it deserves to study the large $N$ index at order
$1$ BPS temperature with this new idea, aiming to find a trace
of the deconfinement phase transition.

Based on this idea, we make a small extension
of \cite{Kinney:2005ej} to probe the deconfinement transition
from the index. More precisely, we find an upper bound of the transition temperature 
by studying the local instability of the confining saddle point. 
Some aspects of this bound is similar to the so-called 
the Hagedorn temperature
\cite{Hagedorn:1965st,Atick:1988si,Aharony:2003sx}.
The similarity arises from the fact that a tachyon condensation instability appears 
to the confining saddle point \cite{Aharony:2003sx}.
The bound we find is indeed order $1$ in the unit of $S^3$ radius,
obtained by optimizing the phases of fugacities. 

We discuss the implication of our bound to the Hawking-Page transition 
of the AdS dual. Curiously, our upper bound turns out to
be lower than the Hawking-Page transition `temperature' of the known analytic black hole
solution of \cite{Gutowski:2004ez}. By the latter, we mean the point at which these
black holes start to dominate over thermal gravitons.
We interpret our finding as predicting new BPS black holes,
with lower transition temperature. It is tempting to conjecture that they are
hairy black holes, similar to those of
\cite{Markeviciute:2018yal,Markeviciute:2018cqs}.

We further sketch a possible scenario on how a first order deconfinement
transition may happen below our bound.
Note that in the partition function of \cite{Aharony:2003sx}, without $(-1)^F$ insertion,
there is a plenty of room for this to happen because the partition function
depends on the coupling constant. Indeed, studying the interaction effects,
\cite{Aharony:2003sx} suggested a mechanism in which a first order deconfinement
transition can happen below the Hagedorn temperature.
In the index, this mechanism cannot be realized since one should trust the
free QFT calculus. We suggest a new mechanism (without any quantitative studies)
of how a deconfinement transition may be
realized below our bound in the index.

The remaining part of this paper is organized as follows. After developing the basic setup
at the beginning of section 2, we compute an upper bound of the deconfinement transition 
temperature from the index in
section 2.1, by optimally tuning the phases of fugacities in the index. In section 2.2,
we revisit the high temperature Cardy-like behavior studied in \cite{Choi:2018hmj}.
In section 2.3, we speculate on how new black hole saddle points would appear in the index
below the Hagedorn-like upper bound. Section 3
concludes the paper with some discussions and remarks.

\section{The large $N$ index at complex fugacities}

The index of 4d $\mathcal{N}=4$ Yang-Mills theory was found in
\cite{Kinney:2005ej,Romelsberger:2005eg}. Its definition is given by
\begin{equation}
  Z(\Delta_I,\omega_i)={\rm Tr}\left[(-1)^Fe^{-\sum_{I=1}^3\Delta_IQ_I
  -\sum_{i=1}^2\omega_iJ_i}\right]\ ,
\end{equation}
with the constraint
\begin{equation}\label{index-constraint}
  \Delta_1+\Delta_2+\Delta_3-\omega_1-\omega_2=0
\end{equation}
on the chemical potentials. $Q_I$ with $I=1,2,3$ are three $U(1)^3\subset SO(6)$
R-charges, and $J_i$ with $i=1,2$ are two $U(1)^2\subset SO(4)$ angular momentum on
spatial $S^3$. They are all normalized so that fermionic fields assume
$\pm\frac{1}{2}$ eigenvalues. This index counts states whose energy is given by
$E=Q_1+Q_2+Q_3+J_1+J_2$, when the $S^3$ radius is multiplied to $E$ to make it
dimensionless. See, e.g. \cite{Choi:2018hmj} for more explanation on our notation.
The free QFT calculus with the $U(N)$ gauge group yields the following
unitary matrix integral form of the index \cite{Kinney:2005ej}:
\begin{equation}\label{index-matrix}
  Z=\frac{1}{N!}\int\prod_{a=1}^N\frac{d\alpha_a}{2\pi}\cdot\prod_{a<b}
  \left(2\sin\frac{\alpha_{ab}}{2}\right)^2\exp\left[\sum_{a,b=1}^N
  \sum_{n=1}^\infty\frac{1}{n}
  \left(1-\frac{\prod_{I=1}^3 2\sinh\frac{n\Delta_I}{2}}
  {2\sinh\frac{n\omega_1}{2}\cdot 2\sinh\frac{n\omega_2}{2}}\right)e^{in\alpha_{ab}}\right]
\end{equation}
where $\alpha_{ab}\equiv\alpha_a-\alpha_b$. $\alpha_a$'s are the $U(1)^N\subset U(N)$
gauge holonomies along the temporal circle, if one interprets this as a partition function
of a Euclidean QFT on $S^3\times S^1$.

As pointed out in \cite{Choi:2018hmj},
we shall give nonzero imaginary parts of $\Delta_I,\omega_i$ compatible with
(\ref{index-constraint}). This will turn out to yield phase factors of fugacities,
obstructing `cancelations' between bosonic/fermionic states at nearby charges.
This schematic idea was already explained in the introduction.
Making a macroscopic saddle point approximation of the inverse Laplace transformation
of the index at charges $\sim N^2$, one wishes to see if one captures macroscopic
entropies. Macroscopic charges are insensitive to whether they are integers or
half-integers. In particular, it is unclear whether the saddle point approximation computes
$+(\textrm{degeneracy})$ or $-(\textrm{degeneracy})$.
Due to a rapid oscillation between $\pm$ signs in the index as one
changes charges by `indistinguishable' units,
the apparent degeneracy captured by the index may look much smaller
than it actually is. Our suggestion is to try to maximally improve this situation by
inserting extra phase factors for fugacities, making the rapid oscillation milder,
or hopefully absent in favorable cases. A priori, we merely try an optimal obstruction
of rapid oscillation, hoping to provide a better lower bound on the true BPS entropy from
the index. In case the lower bound saturates the entropy of known black holes, as
in \cite{Choi:2018hmj}, this approach would count them. However, still we modestly
have the general possibilities in mind: we seek for possible lower bounds for entropies,
which probably will mean upper bounds on various transition temperatures.
Conservatively, most of the results in this paper in principle has to be interpreted
this way. However, such bounds will lead to interesting predictions on the gravity duals.

Once we complexify the chemical potentials $\Delta_I,\omega_i$, the effective potential
for $\alpha_a$ appearing in (\ref{index-matrix}) (minus log of the integrand)
will be complexified. Then the large
$N$ saddle points for $\alpha_a$ may deviate from real $\alpha_a$, i.e. away from the
unit circle in the space of $e^{i\alpha_a}$. Finding the large $N$ saddle points in this
complex plane appears to be a difficult problem.
We shall discuss it only briefly in section 2.3. Here, we first review the large $N$ analysis
of the index at real fugacities \cite{Kinney:2005ej}, where the saddle points for
$e^{i\alpha_a}$ all stay on the unit circle, and slightly improve it in section 2.1 to
see a tachyon instability from the index.

\cite{Kinney:2005ej,Aharony:2003sx} replaces the integrals over
a large number of variables $\alpha_a$ by a functional integral over the distribution
function $\rho(\theta)$ of $N$ particles on a circle. Here, $\theta\sim\theta+2\pi$.
The exact, or fine-grained, distribution for $N$ particles would have been
\begin{equation}\label{rho-exact}
  \rho(\theta)=\frac{1}{N}\sum_{a=1}^N\delta(\theta-\alpha_a)=
  \frac{1}{2\pi N}\sum_{n=-\infty}^\infty\sum_{a=1}^Ne^{in(\theta-\alpha_a)}\ ,
\end{equation}
with the normalization $\int_0^{2\pi}d\theta \rho(\theta)=1 $.
At large $N$, with a dense distribution of eigenvalues along the circle,
we coarse-grain $\rho(\theta)$ to generic functions.
One may Fourier expand $\rho(\theta)$ as
\begin{equation}\label{rho-fourier}
  \rho(\theta)=\frac{1}{2\pi}+
  \frac{1}{2\pi}\sum_{n=1}^\infty\left[\rho_ne^{in\theta}+\rho_{-n}e^{-in\theta}\right]
  \ \ \ ,\ \ \ \rho_{-n}=\rho_n^\ast\ .
\end{equation}
This function is subject to the local constraint $\rho(\theta)\geq 0$. The global constraint
$\int_0^{2\pi} d\theta\rho(\theta)=1$ is already solved in the above expression.
In the exact fine-grained expression (\ref{rho-exact}), the $n$'th Fourier coefficient
$\rho_n$ is given by
\begin{equation}\label{fourier-exact}
  \rho_n=\frac{1}{N}\sum_{a=1}^Ne^{-in\alpha_a}\ .
\end{equation}
The functional integral form of $Z$ in the
large $N$ limit is given by \cite{Kinney:2005ej}
\begin{equation}
  Z=\int\prod_{n=1}^\infty\left[d\rho_n d\rho_{-n}\right]
  \exp\left[-N^2\sum_{n=1}^\infty\frac{1}{n}\rho_n\rho_{-n}
  \frac{\prod_I(1-e^{-n\Delta_I})}{\prod_{i}(1-e^{-n\omega_i})}\right]\ .
\end{equation}
Here, in the manipulation, we used $\sum_I\Delta_I=\sum_i\omega_i$.

For simplicity, from now on, let us consider the case with equal charges,
$Q_1=Q_2=Q_3\equiv Q$, $J_1=J_2\equiv J$.
Then one sets $\Delta_1=\Delta_2=\Delta_3\equiv \Delta$,
$\omega_1=\omega_2\equiv\omega$, satisfying
$3\Delta=2\omega$. We label $e^{-\omega}=x^3$, $e^{-\Delta}=x^2$.
Then one finds
\begin{equation}\label{index-large-N}
  Z=\int\prod_{n=1}^\infty\left[d\rho_n d\rho_{-n}\right]
  \exp\left[-N^2\sum_{n=1}^\infty\frac{f(x^n)}{n}\rho_n\rho_{-n}\right]
\end{equation}
with
\begin{equation}
  f(x)=\frac{(1-x^2)^3}{(1-x^3)^2}\ .
\end{equation}
At real fugacity in the physical range $0<x<1$, $f$ is positive. This implies that all the mode
integrals over $\rho_n$ can be approximated by a Gaussian integral around $\rho_n=0$.
Since the large $N$ saddle point is a uniform distribution $\rho(\theta)=\frac{1}{2\pi}$,
one does not have to worry about the positivity constraint $\rho(\theta)\geq 0$.
The resulting partition function is given by
\begin{equation}\label{graviton-index}
  Z\sim\prod_{n=1}^\infty f(x^n)^{-1}=\prod_{n=1}^\infty\frac{(1-x^{3n})^2}{(1-x^{2n})^3}\ ,
\end{equation}
and agrees with the index over gravitons in $AdS_5\times S^5$ \cite{Kinney:2005ej}.
(This analysis was done in \cite{Kinney:2005ej} with all $4$ fugacities kept.)
Since the free energy is independent of $N$, the index does not see deconfinement
at arbitrary high `temperature' (meaning $x$ close to $1$, or $\omega$ close to $0$).

On the other hand, in the partition function without $(-1)^F$ at weak coupling,
the term $f(x^n)$ appearing in (\ref{index-large-N}) is replaced by \cite{Aharony:2003sx}
\begin{equation}
  1-z_{\rm B}(x^n)-(-1)^{n-1}z_F(x^n)\ .
\end{equation}
$z_B$ and $z_F$ are bosonic and fermionic parts of the `letter partition function'
respectively. This expression turns negative beyond certain values of $x$,
say at $x>x_H$ for $n=1$. It turns out that the coefficient for $n=1$ becomes negative
first, driving $\rho_1$ to condense. As discussed in \cite{Aharony:2003sx},
this implies that the low temperature saddle point with $\rho_n=0$, preserving
the `winding number symmetry' in the Euclidean picture, seize to exist. So one
identifies $T_H\equiv-\log x_H$ as the Hagedorn temperature of this system.
The actual phase transition to the high temperature deconfining phase may happen
below this temperature, and various scenarios at weak but nonzero coupling are
discussed in \cite{Aharony:2003sx}. In any scenarios, $T_H$ is the upper bound for
the temperature for which the free energy of the dominant saddle point can be
at $\mathcal{O}(N^0)$ order. This allows us to identify $T_H$ as
an upper bound for the deconfinement transition temperature.

\subsection{Instability of the confining saddle point}

Now we introduce a phase for $x$, shifting $x\rightarrow xe^{i\phi}$ with real
$x$, $\phi\sim\phi+2\pi$, and redo the analysis starting from (\ref{index-large-N}).
Now with the complexified effective action, one should allow
$e^{i\alpha_a}$'s away from the unit circle at the saddle points.
This would mean that one will have to generalize the ansatz from the unit
circle to a more general curve on the complex plane. This apparently complicated
task will not be discussed here.

We restrict our interest to the fate of the graviton saddle point, focussing on
the local fluctuations. In (\ref{index-large-N}), we are simply asking whether
the effective action
\begin{equation}\label{effective-action}
  S_{\rm eff}=N^2\sum_{n=1}^\infty\frac{f(x^n)}{n}\rho_n\rho_{-n}
\end{equation}
is locally stable or not around $\rho_n=0$. Clearly, even
with complex $f(x^n)$, $\rho_n=0$ will continue to be an extremum under their small
variations. One simply has to make sure if the real part of $S_{\rm eff}$ is at
its local minimum, and if the imaginary part of it is stationary. If both of
these conditions are met, the Gaussian integration of the virtually unconstrained
small fluctuations $\delta\rho_n$ (around $\rho_n=0$) clearly yields
the known graviton index on $AdS_5\times S^5$ \cite{Kinney:2005ej}, simply with
complexified fugacities.

\begin{figure}[t!]
\begin{center}
	\includegraphics[width=8cm]{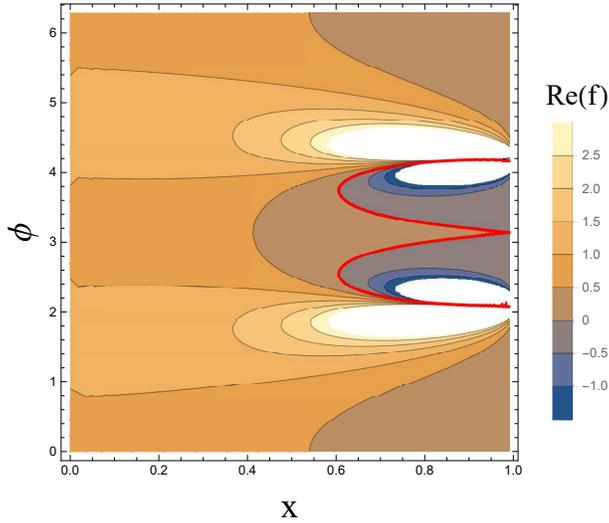}
\caption{Contour plot of ${\rm Re}(f)$ on the $x$-$\phi$ space.
The red line shows the curve ${\rm Re}(f)=0$.}
	\label{deconfine}
\end{center}
\end{figure}

The above analysis will hold if ${\rm Re}(f(x^ne^{in\phi}))>0$.
If this can go negative at finite $x<1$, at optimally tuned $\phi$,
this will imply the disappearance of the graviton saddle point. One should
tune $\phi$ so that ${\rm Re}(f)$ becomes $0$ at lowest possible $x$.
This is because, with boson/fermion cancelation, we see less spectrum and
the phase transitions apparently look delayed or even become invisible in the index.
With minimized boson/fermion cancelations, we can probably see a transition
with minimized delay. So we identify the lowest $x$ with
${\rm Re}(f)=0$ as the `temperature' where tachyon condensation starts. 
We call this value $x_H$.

One finds that ${\rm Re}(f(xe^{i\phi}))$ as a function of $x,\phi$ is given by
\begin{equation}
\frac{(1-x^2) (1 + x^2 -
    2 x \cos\phi)^2 \left(2 x (2 + 5 x^2 + 2 x^4) \cos\phi + (1 +
       x^2) (1 + 4 x^2 + x^4 + 3 x^2 \cos(2 \phi))\right)}{(1 + x^6 -
   2 x^3 \cos(3 \phi))^2}\ .
\end{equation}
All other factors are positive except the last factor on the numerator.
The vanishing condition
\begin{equation}
  2 x (2 + 5 x^2 + 2 x^4) \cos\phi + (1 +
       x^2) (1 + 4 x^2 + x^4 + 3 x^2 \cos(2 \phi))=0
\end{equation}
is solved by
\begin{equation}
  \cos\phi=\frac{-2-5x^2-2x^4\pm\sqrt{-2+2x^2+9x^4+2x^6-2x^8}}
  {6x(1+x^2)}\ .
\end{equation}
This line on the $x$-$\phi$ plane is shown in by Fig. \ref{deconfine} by the red curve.
On the right sides of this curve, one finds ${\rm Re}(f)<0$. In the remaining region,
${\rm Re}(f)>0$.

On the red curve, the minimal value of $x$ (maximal value of
chemical potential $\omega$, meaning minimal `temperature') is
obtained when the two solutions for $\phi$ get degenerate, i.e. when
\begin{equation}
  -2+2x^2+9x^4+2x^6-2x^8=0\ .
\end{equation}
The relevant solutions is $x_H=\sqrt{\frac{\sqrt{3}-1}{2}}\approx 0.605$. This is the point
at which one can optimally tune $\phi$ to trigger the tachyon condensation
at lowest $x$. The tuned value of $\phi$ is given by $\cos\phi=-\frac{1}{2x_H}$, i.e.
$\phi\approx 0.81\pi$ or $\approx(2-0.81)\pi$. The two values of $\phi$'s are symmetric
around $\phi=\pi$, as is manifest from Fig. \ref{deconfine}.
They are at the top of the two dome regions for ${\rm Re}(f)\leq 0$.
This will set the upper bound on the actual deconfinement
transition temperature. At these points, one finds
\begin{equation}
  \omega_H=-3\log x_H\approx 1.508.
\end{equation}
This is higher than the Hawking-Page transition point
\begin{equation}
  \omega_{\rm HP}^{\rm known}=\frac{\pi}{16}\sqrt{414-66\sqrt{33}}\approx 1.159
\end{equation}
of the known black holes, computed in section 2.3 of \cite{Choi:2018hmj}.
See also our section 2.3 below for a review and summary.
Our upper bound $\omega_H^{-1}$ is lower than the
Hawking-Page temperature of known black holes,
$\omega_H^{-1}<(\omega_{\rm HP}^{\rm known})^{-1}$.

Let us think about the implications of this finding. 
We have found the temperature $\omega_H^{-1}$ where the confining saddle point 
would have a local instability. As we increase the temperature, the system should 
transit to a new phase at or before this point. This transition would be 
a Hawking-Page transition. But since
$\omega_H^{-1}<(\omega_{\rm HP}^{\rm known})^{-1}$, the transition cannot 
be realized by the known black hole solutions.
So this naturally indicates the existence of new, yet
undiscovered, BPS black holes in $AdS_5\times S^5$. If these hypothetical
black holes have lower Hawking-Page temperature than $\omega_H^{-1}$, they will dominate
over thermal gravitons below the bound we computed.

In fact, numerical solutions for (almost) BPS black holes are found
\cite{Markeviciute:2018yal,Markeviciute:2018cqs} in the sector we studied,
$Q_1=Q_2=Q_3\equiv Q$, $J_1=J_2\equiv J$. However, their charges $Q,J$ seem to
be too small to be relevant for this transition. See section 2.3 for more discussions
on small and large BPS black holes. Also, it is not a priori clear whether the
consistent truncations used to construct these solutions would capture the most dominant
saddle points. In any case, we find it a very promising signal that more general BPS
black holes than those of \cite{Gutowski:2004ez,Gutowski:2004yv,Chong:2005da,Kunduri:2006ek}
are being found.

The tachyon instability of $\rho_1$ has some similarities with 
the Hagedorn behavior in the partition function of \cite{Aharony:2003sx}. In particular, 
as one approaches $x\rightarrow x_H$ from below, the density of states exhibits an 
exponential growth \cite{Aharony:2003sx}. However, in the index, this feature is 
not visible in the graviton index (\ref{graviton-index}). Namely, due to nonzero 
${\rm Im}(f)$ at $x_H$, $\cos\theta=-\frac{1}{2x_H}$, the index remains finite 
even at $x=x_H$.

For $x>x_H$, $\rho_1$ should condense. The free energy is expected
to be of order $N^2$. In this regime, $\omega<\omega_H$, there seem to be no reason
to expect that the true saddle point for $e^{i\alpha_a}$'s
be on the unit circle. So it seems that we cannot apply the studies made in
\cite{Aharony:2003sx}, beyond the transition.

At $x<x_H$, whether the saddle point with $\rho_n=0$ is a global one or not is of course
unclear. To this end, one should make a more global study, again at more general
contour on the space of $e^{i\alpha_a}$. We only comment on it briefly in section 2.3.

\subsection{Cardy limit revisited}

Despite the complication stated at the end of section 2.1, due to complex effective
action, one can still make a quantitative analysis at $\omega=-3\log x\ll 1$.
(Here, $x$ means the real modulus of the complex fugacity $xe^{i\phi}$.)
This is the so-called `Cardy limit' studied in \cite{Choi:2018hmj}.
To see this, consider the following 2-body potential
\begin{eqnarray}\label{Veff}
  V_{\rm eff}(\theta)&=&-\log\left(2\sin\frac{\theta}{2}\right)^2
  +\sum_{n=1}^\infty\frac{1}{n}\left(f(x^ne^{in\phi})-1\right)(e^{in\theta}+e^{-in\theta})\\
  &=&-\log\left(2\sin\frac{\theta}{2}\right)^2
  +\sum_{n=1}^\infty\frac{1}{n}\left(\frac{(1-x^{2n}e^{2in\phi})^3}
  {(1-x^{3n}e^{3in\phi})^2}-1\right)(e^{in\theta}+e^{-in\theta})\nonumber
\end{eqnarray}
between two eigenvalues $\alpha_a$, $\alpha_b$, where $\theta=\alpha_{ab}$.
This leads to a `force' on the complex $\theta$ plane, which is in fact a cylinder with
$\theta\sim\theta+2\pi$, given by
\begin{equation}\label{force}
  -\frac{\partial V_{\rm eff}}{\partial\theta}=
  \cot\frac{\theta}{2}+2\sum_{n=1}^\infty\left(\frac{(1-x^{2n}e^{2in\phi})^3}
  {(1-x^{3n}e^{3in\phi})^2}
  -1\right)\sin (n\theta)\ .
\end{equation}
The first term coming from the Haar measure behaves like $\sim\frac{2}{\theta}$ at small
$\theta$, which is repulsive at real $\theta$. Had $\theta$ been real and nonzero
(even if small), one could have
rearranged part of the second term in $V_{\rm eff}$ as
\begin{equation}
  -\sum_{n=1}^\infty\frac{1}{n}(e^{in\theta}+e^{-in\theta})=
  \log(1-e^{i\theta})(1-e^{-i\theta})=\log\left(2\sin\frac{\theta}{2}\right)^2\ ,
\end{equation}
canceling the first term of $V_{\rm eff}$. However, for complex $\theta$,
separating terms in the sum over $n$ could be dangerous.

Now let us consider the second term of $V_{\rm eff}$ in the `high temperature limit' $\omega\rightarrow 0^+$. In the index, this limit may or may not be nontrivial, depending
on the value of $\phi$. For instance, at $\phi=0$ and $0<x<1$, the index will never
exhibit a macroscopic entropy as shown in \cite{Kinney:2005ej}. The crucial reason
for this was that ${\rm Re}(f(x^n))$ remained positive, as shown in Fig. \ref{deconfine}
along the $x$-axis. However, note that beyond $x>x_H=\sqrt{\frac{\sqrt{3}-1}{2}}$,
there is a region in the $x$-$\phi$ plane which has ${\rm Re}(f)<0$, providing chances
for a macroscopic entropy. Even though the analysis of section 2.1 was limited to the
situation where $e^{i\alpha_a}$'s sit on the unit circle, it is still an important question
whether ${\rm Re}(f(x^ne^{in\phi}))$ can go negative, since this will allow
$V_{\rm eff}(\theta)$ to have negative real part even at (small) complex $\theta$.
So we carefully re-investigate the results of section 2.1 on the behaviors of
${\rm Re}(f(x^ne^{in\phi}))$.

We first study the term with $n=1$, i.e. ${\rm Re}(f(xe^{i\phi}))$. It will turn out
that understanding this term will be most important even in the Cardy limit.
The region with ${\rm Re}(f(xe^{i\phi}))<0$ is on the right side of the red curve shown in
Fig. \ref{deconfine}, consisting of the `dome' regions.
Therefore, if one wishes to take the Cardy limit
$x\rightarrow 1^-$, one should again keep $\phi$ at an optimal value in this region,
to maximally obstruct cancelations of nearby bosons/fermions.
For the term with $n=1$, it is easy to see from Fig. \ref{deconfine} how to set $\phi$,
as $x\rightarrow 1^-$.
This is easily noticed by following the valley of lowest ${\rm Re}(f)$ inside
the dome. At $x=x_H$, the optimal value was shown to be
$\phi=\cos^{-1}\left(-\frac{1}{2x_H}\right)\approx 0.81\pi$. From here, we only
consider the lower dome, $\phi\leq \pi$.
As one further increases $x$, the value of $\phi$ which minimizes
${\rm Re}(f(xe^{i\phi}))$ will decrease, towards $\phi\searrow\frac{2\pi}{3}$
as $x\rightarrow 1^-$. Namely, setting $\phi=\frac{2\pi}{3}$, ${\rm Re}(f(xe^{i\phi}))$
will maximally diverge to $-\infty$ as $x\rightarrow 1^-$.

We would like to see this behavior more quantitatively, including all other terms
at higher $n$'s in $V_{\rm eff}$. Let us take $x=e^{-\frac{\omega}{3}}$ with
$\omega\ll 1$ and $\phi\approx\frac{2\pi}{3}$. Then one finds
\begin{equation}\label{f-cardy}
  \frac{(1-x^{2n}e^{2in\phi})^3}{(1-x^{3n}e^{3in\phi})^2}\approx
  \frac{(1-x^{2n}e^{\frac{4\pi n i}{3}})^3}{(1-x^{3n})^2}\approx
  \frac{1}{n^2\omega^2}(1-e^{\frac{4\pi ni}{3}})^3\ .
\end{equation}
At $n\neq 1$, the real part of this term will oscillate in its sign. Therefore,
it may not be clear at this stage whether setting $\phi=\frac{2\pi}{3}$ is an ideal one
or not. A more general study can be made by setting $\phi$ to be an arbitrary real number
between $0$ and $2\pi$, and maximize $\log Z$ or the entropy after all the calculus.
This was in fact done in \cite{Choi:2018hmj} (with maximally deconfining distribution,
to be addressed shortly below), which indeed confirms that
$\phi=\frac{2\pi}{3}$ is the optimal one. So with this understood, we shall set
$\phi=\frac{2\pi}{3}$ in this paper for the simplicity of presentation.

Since this term (\ref{f-cardy}) is dominant in (\ref{force})
due to the diverging factor $\frac{1}{\omega^2}$, the vanishing force condition
at the leading order requires $\sum_{n=1}^\infty
\frac{(1-e^{\frac{4\pi in}{3}})^3}{n^2\omega^2}\sin(n\theta)\approx 0$.\footnote{A more
careful treatment of the sum over $n$, separating $n\lesssim|\omega|^{-1}$ and
$n\gtrsim|\omega|^{-1}$, was presented in \cite{Choi:2018hmj}.}
So the leading order solution at small $\omega$ is $\theta\approx 0$ for all pairs
$\alpha_a,\alpha_b$, i.e. the maximally deconfining configuration.
Since all matters are in the adjoint representation, it does not matter in the leading
order in $\omega$ whether $e^{i\alpha_a}$'s stay on the unit circle or not.
These are precisely the Cardy saddle points considered in \cite{Choi:2018hmj}.
As in \cite{Choi:2018hmj}, we assume the global dominance of this saddle point.

With the discussions in the previous paragraph,
we can regard the eigenvalues as asymptotically living on the unit circle.
Thus we can use the formula (\ref{effective-action}),
where $\rho_n$ are Fourier coefficients of the distribution on unit circle.
Just like the studies made in section 5.3 of \cite{Aharony:2003sx} for
the maximally deconfining saddle point, we set $\rho_n=1$ for $\rho(\theta)=\delta(\theta)$.
One thus obtains
\begin{equation}\label{cardy-intermediate}
  \log Z\sim -S_{\rm eff}=-N^2\sum_{n=1}^\infty\frac{f(x^n)}{n}\rho_n\rho_{-n}
  \approx-\frac{N^2}{\omega^2}\sum_{n=1}^\infty\frac{(1-e^{\frac{4\pi in}{3}})^3}{n^3}
  =\frac{3N^2}{\omega^2}\left({\rm Li}_3(e^{\frac{4\pi i}{3}})
  -{\rm Li_3}(e^{\frac{8\pi i}{3}})\right)\ .
\end{equation}
${\rm Li}_3(z)=\sum_{n=1}^\infty\frac{z^n}{n^3}$ converges for
$|z|<1$, and also at $|z|=1$ if $z\neq 1$ (i.e. not at the branch point of this function).
Here, note that
\begin{equation}\label{Li3-identity}
  {\rm Li}_3(e^{\frac{4\pi i}{3}})-{\rm Li_3}(e^{\frac{8\pi i}{3}})=
  \frac{1}{6}\left(\frac{2\pi i}{3}\right)^3\ .
\end{equation}
This can be proved by using an identity of ${\rm Li}_3$ and the Bernoulli
polynomial $B_3$, as in \cite{Choi:2018hmj}. Alternatively, one can confirm this
simply by performing the infinite sums on the left hand side. For instance,
as a brutal but clearest check, we reconfirmed it numerically by computing
the infinite sum till $n=1000$, finding that both sides are
$\approx -1.53117 i$. So one obtains
\begin{equation}\label{Cardy-special}
  \log Z\sim\frac{N^2\left(\frac{2\pi i}{3}\right)^3}{2\omega^2}\ ,
\end{equation}
at $\omega\ll 1$. This is the specialization of the Cardy-like formula found in
\cite{Hosseini:2017mds,Choi:2018hmj},
\begin{equation}\label{Cardy-general}
  \log Z\sim\frac{N^2\Delta_1\Delta_2\Delta_3}{2\omega_1\omega_2}\ \ ,\ \ \
  \Delta_1+\Delta_2+\Delta_3-\omega_1-\omega_2=2\pi i\ .
\end{equation}
Restricting to the case $\Delta_1=\Delta_2=\Delta_3\equiv\Delta$ and
$\omega_1=\omega_2\equiv\omega\ll 1$, one obtains $\Delta\approx\frac{2\pi i}{3}$.
So (\ref{Cardy-general}) indeed reduces to (\ref{Cardy-special}) in the setting of
this subsection.

\subsection{Comments on deconfinement}

We shall discuss a possible mechanism, or a scenario, on
how deconfinement transition may happen below our bound computed in section 2.1. 
This subsection will be rather speculative.
Before presenting our speculations, we first discuss the possible properties of
the black hole saddle points, by considering the known black hole solutions.

It is perhaps illustrative to start the discussions from the well-known
AdS Schwarzschild black holes. The AdS$_5$ Schwarzschild black holes have mass
(energy) $M$, conjugate to the temperature $T$. The relations of $M$, $T$ and the
horizon radius $r_+$ is given by
\begin{equation}\label{AdS-Sch}
  T=\frac{r_+}{\pi\ell^2}+\frac{1}{2\pi r_+}\ \ ,\ \ \
  r_+^2=-\frac{\ell^2}{2}+\ell\sqrt{\frac{\ell^2}{4}+\omega M}\ ,
\end{equation}
where $\ell$ is the radius of AdS$_5$, and
$\omega\equiv\frac{16\pi G_N}{3{\rm vol}(S^3)}$ with 5d Newton constant
$G_N$, and ${\rm vol}(S^3)$ is the volume of unit 3-sphere.
For instance, see \cite{Witten:1998zw} for its summary.
$r_+$ is a monotonically increasing function of $M$, and thus labels the
energy to certain extent. From the expression of $T$, one finds that the black hole
saddle points exist only at $T\geq T_0\equiv\frac{\sqrt{2}}{\pi\ell}$.
At given temperature $T>T_0$, two black hole solutions exist, solving the first equation
of (\ref{AdS-Sch}). The one with smaller $r_+$ has negative specific heat,
$\frac{\partial r_+(T)}{\partial T}<0$ and thus $\frac{\partial M}{\partial T}<0$,
irrelevant for discussing canonical ensemble. The solution with larger $r_+$ is
called large black holes, having positive specific heat.

One should discuss the thermodynamics with two saddle points: large black holes
and thermal gravitons in AdS$_5$. The thermal graviton phase is dominant
at $T<T_{\rm HP}$ with $T_{\rm HP}=\frac{3}{2\pi\ell}$, while
the large black hole is dominant at $T>T_{\rm HP}$
\cite{Hawking:1982dh,Witten:1998zw,Aharony:2003sx}. Since the free energy of thermal
gravitons is of order $\mathcal{O}(N^0)$ in the large $N$ limit while that of
the black hole is $\mathcal{O}(N^2)$, the dominant saddle point is determined by
the sign of the black hole free energy.
The transition is known to
be of first order, called Hawking-Page transition. The gauge theory dual picture of
this transition is the confinement-deconfinement transition at strong coupling
\cite{Witten:1998zw,Aharony:2003sx}.

To summarize, some characteristic properties of this system are:
(1) the appearance of a local saddle point at $T=T_0$, below the transition temperature
$T_{\rm HP}$; (2) having two branches of black holes, where the small black hole is meaningful
only in the micro-canonical ensemble and contains the small charge limit; (3) the
transition happens while the `graviton saddle point' locally exists.

\begin{figure}[t!]
\begin{center}
	\includegraphics[width=8cm]{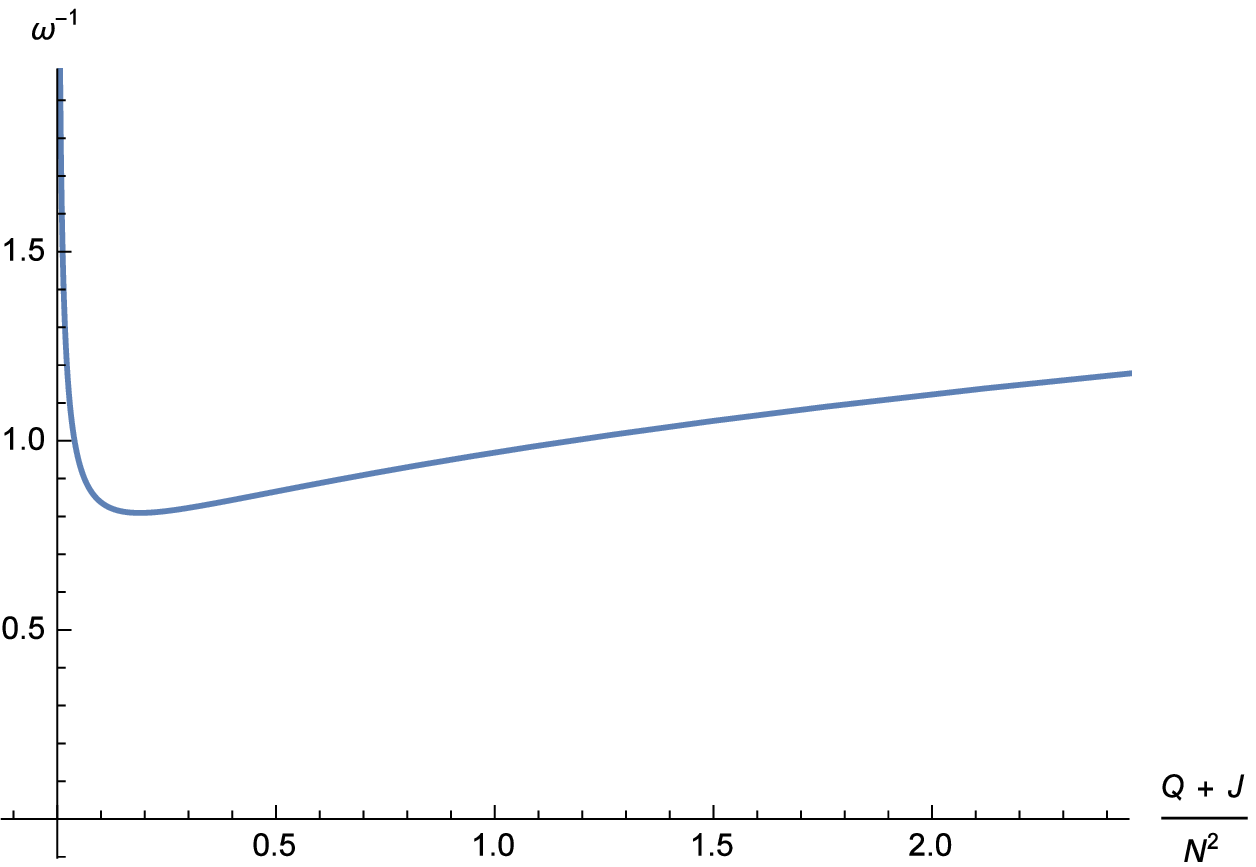}\hspace{.7cm}
    \includegraphics[width=8cm]{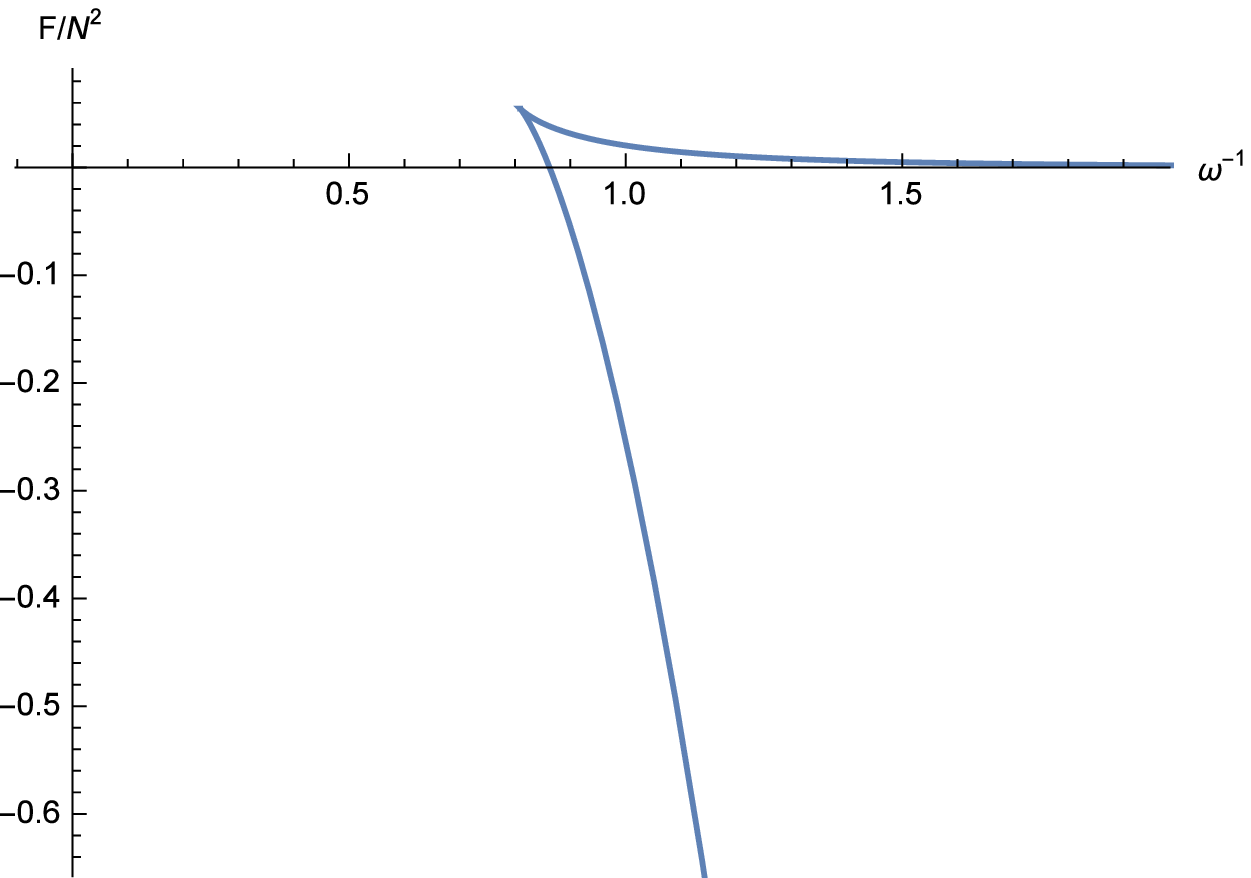}
\caption{[left] Charge vs. temperature. There are small and large
black hole branches, with negative/positive specific heat, respectively.
[right] Temperature vs. free energy. The upper curve for is small black holes with
positive free energy, always losing against thermal AdS gravitons. The lower curve is for
large black holes, dominating for $\omega<\omega_{\rm HP}^{\rm known}$ with $F<0$.}
	\label{BH-thermo}
\end{center}
\end{figure}

We move on to the BPS thermodynamics, for the solutions
known in the literature. We focus on the case with
$Q\equiv Q_1=Q_2=Q_3$, $J\equiv J_1=J_2$, and study the BPS states at
$\Delta E\equiv E-3Q-2J=0$. The last projection implies taking the
Hawking temperature of black holes to zero. In this BPS sector,
the positive charges $Q$, $J$ contribute to the BPS energy as
$3Q+2J$. The fugacity $x=e^{-\frac{\omega}{3}}$ introduced in
sections 2.1 and 2.2 couple to $Q+J$, like
${\rm Tr}\left[e^{-2\omega(Q+J)}\right]$ in the partition function.
We shall consider the `BPS thermodynamics'
of the analytic black holes solutions of \cite{Gutowski:2004ez}.
It is shown in \cite{Hosseini:2017mds,Choi:2018hmj} that the entropy and the chemical
potential $\omega$ of these black holes can be computed
by making a Legendre transformation of
\begin{equation}\label{free-known-BH}
  \log Z\sim\frac{N^2\left(\frac{2\pi i}{3}+\frac{2}{3}\Omega\right)^3}{2\Omega^2}\ .
\end{equation}
Namely, one extremizes the following entropy function
\begin{equation}
  S(\Omega;Q+J)=\log Z+2\Omega(Q+J)
\end{equation}
in $\Omega$. The entropies  of the known black holes of \cite{Gutowski:2004ez} are
reproduced by taking ${\rm Re}(S)$ of the extremized $S(\Omega;Q+J)$ in $\Omega$
\cite{Hosseini:2017mds}, and the chemical potential $\omega$ is obtained by the extremal
value of ${\rm Re}(\Omega)$ \cite{Choi:2018hmj}. Strictly speaking, these agreements
are checked
by applying a charge relation met by $Q$ and $J$ in the known solution of \cite{Gutowski:2004ez}.
With this understood, The real part of (\ref{free-known-BH}) is minus of
the free energy of the known black holes of \cite{Gutowski:2004ez}.
We summarize the extremal values of $\omega$, the BPS free energy
$F=-{\rm Re}(\log Z)$, and the entropy ${\rm Re(S)}$, all worked out in detail
in section 2.3 of \cite{Choi:2018hmj}. One finds
\begin{eqnarray}
  \omega&=&-\xi\sqrt{\frac{3\pi+3\xi}{\pi-3\xi}}
  \ \ ,\ \ -\pi<\xi<0\nonumber\\
  F&=&-{\rm Re}(\log Z)=-\frac{N^2}{18}\frac{\pi^3-9\pi\xi^2-8\xi^3}{\xi^2}
  \sqrt{\frac{\pi+\xi}{3\pi-9\xi}}\nonumber\\
  Q+J&=&-\frac{N^2}{54}\frac{(\pi-2\xi)^2(\pi+\xi)}{\xi^3}\ .
\end{eqnarray}
The plots for the `temperature' $\omega^{-1}$, $\frac{F}{N^2}$, charge
$\frac{Q+J}{N^2}$ are shown in Fig.
\ref{BH-thermo}. Let us call $T\equiv\omega^{-1}$ the `temperature' as this plays this
role, conjugate to $Q+J$. From the left figure, one finds that there are two branches
of black holes for $T>T_0\equiv\left[\pi\sqrt{\frac{2}{\sqrt{3}}-1}\right]^{-1}
\approx 1.24^{-1}$,  similar to the AdS-Schwarzschild black holes. In the small
black hole branch, the specific heat (the slope) is negative. So we do not consider this
saddle point if we are in the grand canonical ensemble. The large black hole branch is to compete
with the thermal BPS graviton phase, at $\frac{F}{N^2}\approx 0$. From the graph on
the right side of Fig. \ref{BH-thermo}, one finds that the large black hole dominates over
thermal BPS gravitons for
\begin{equation}
  T^{-1}=\omega<\omega_{\rm HP}^{\rm known}
  \equiv\frac{\pi}{16}\sqrt{414-66\sqrt{33}}\approx 1.16\ ,
\end{equation}
which corresponds to $Q+J>\frac{3+\sqrt{33}}{18}N^2\approx 0.486N^2$.

As advertised in section 2.1, the would-be Hawking-Page transition temperature
$(\omega_{\rm HP}^{\rm known})^{-1}$ of these black holes is higher than our
upper bound $\omega_{\rm H}^{-1}\approx 1.508^{-1}$.
Therefore, had this black hole been the only BPS black holes in $AdS_5\times S^5$,
one would have arrived at a contradiction. As stated before, a natural rescue seems to be
the existence of yet unknown BPS black hole saddle points with lower transition temperature.
Although these conjectured black holes appear to be in the same charge sector as the studied
hairy BPS black hole solutions \cite{Markeviciute:2018yal,Markeviciute:2018cqs}, their
charges seem to be too small to cover the black holes at order $1$ `temperature.'

The analytic BPS black hole solutions explained so far
are somewhat similar to the AdS-Schwarzschild black holes.
They have two branches, for small and large black holes. The small black hole
branch has negative specific heat, $\frac{\partial(Q+J)}{\partial(\omega^{-1})}<0$.
It is not clear whether the yet unknown black hole solutions that we claim
also have such structures.
However, we feel it desirable to seek for possible local saddle points appearing at a
temperature $T_0$ below our $T_H=\omega_H^{-1}$, which overtakes
the thermal graviton saddle point at a higher temperature by a first order phase
transition, still below $T_H$ that we computed.

In \cite{Aharony:2003sx}, the large $N$ partition function (without $(-1)^F$)
was studied at
weak but nonzero coupling, addressing a possible scenario for a first order phase transition
below the Hagedorn temperature. It was crucial that the partition function depends
on the coupling constant to realize a first order deconfinement phase transition below
the Hagedorn temperature. However, in the BPS sector, we are now studying an index which
is supposed to be independent of the coupling constant (unless very drastic situations happen).

It seems that a possible mechanism for a first order phase transition below $T_H$
in the index (if this is the right scenario at all) is the complexified saddle points
of $e^{i\alpha_a}$'s away from the unit circle. We have been emphasizing this
possibility throughout this paper. But due to our technical limitation, we only
studied the saddle point on the unit circle satisfying $\rho_n=0$, 
thus only being able
to study its local instabilities. It may be possible that a disconnected complex
saddle point in the $e^{i\alpha_a}$ plane may suddenly appear at a temperature
$T_0<T_H$, then perhaps branching into small/large black holes. At the very least,
the known black hole saddle points of \cite{Gutowski:2004ez} should be identified,
although they are likely to be subdominant around their creation till around
$(\omega_{\rm HP}^{\rm known})^{-1}$. (They are likely to be the dominant saddle points
in the large charge limit \cite{Choi:2018hmj}.)

We would like to further seek for such new saddle points. In general, one should
solve integral equations containing the curve $r(\theta)$, where the distribution of
$e^{i\alpha_a}$'s is labeled by the radius $r$ at a given angle $\theta$,
and the distribution $\rho(\theta)$ along the curve. For instance, this was solved
in the partition function of 3d SCFT's on M2-branes on $S^3$ \cite{Drukker:2010nc,Herzog:2010hf}.
Compared to these works, our problem appears to be much more difficult in the following
sense. In our effective potential $V_{\rm eff}$, say in (\ref{Veff}), there are infinitely
many terms in $f(x)$ in the Taylor expansion in $x$, since $f$ is a rational function rather
than a polynomial. The problem of large $N$ eigenvalue distributions of
\cite{Herzog:2010hf} is like replacing the infinite series $f(x)$ in $x$ by a finite polynomial.
Perhaps it will be easier to find the first nucleation of such saddle points at $T=T_0$,
when two branches will be degenerate so that a further non-integral saddle point condition
can be imposed. We would like to come back to this problem hopefully in a near future.

\section{Discussions}

In this paper, we pointed out that the index of $\mathcal{N}=4$ Yang-Mills theory
on $S^3\times\mathbb{R}$ should undergo a large $N$ phase transition. 
A key idea is to turn on the finite phases of BPS fugacities,
to optimally obstruct boson/fermion cancelations of nearby BPS states at macroscopic charges.
We compute a temperature which sets an upper bound of the confinement-deconfinement
transition of the gauge theory in the BPS sector, or equivalently the Hawking-Page
transition of BPS black holes in AdS.

One would hope to better understand the actual transition
from the index. We think our calculations and arguments clearly
indicate the existence of such a transition, visible in the index.
Unfortunately, the large $N$ saddle point analysis of the index appears
technically tricky, and we leave this interesting question for future studies.
However, the studies of this paper and of \cite{Choi:2018hmj} shed concrete lights
on the  BPS black holes in $AdS_5\times S^5$.
Among others, there are signals that new black holes have to exist,
beyond the analytic solutions found in
\cite{Gutowski:2004ez,Gutowski:2004yv,Chong:2005da,Kunduri:2006ek}.
Considering the studies made in \cite{Markeviciute:2018yal,Markeviciute:2018cqs},
it is tempting to conjecture that the new black holes
are hairy black holes. Unfortunately, the numerical solutions in
\cite{Markeviciute:2018yal,Markeviciute:2018cqs} are found only for small charges
$Q,J\ll N^2$.\footnote{We thank Jorge Santos for explaining to us some properties
of known numerical solutions.} Although it may sound technically challenging, one would
still like to ask whether one can find hairy BPS black holes in $AdS_5\times S^5$
at charges of order $N^2$.

Turning the logic around, one would also like to find
(perhaps unstable) saddle points of the large $N$ index at small charges, to
study small AdS black holes in the micro-canonical ensemble.
For instance, it will be interesting to see if the non-interacting mix picture
\cite{Bhattacharyya:2010yg} between the small black hole and the hair can be
confirmed from the QFT side. See also \cite{Markeviciute:2016ivy}.

More generally, it will be desirable to further study how rich the landscape of BPS
black holes is in $AdS_5\times S^5$. It is almost certain to us that BPS hairy black holes
will be playing prominent roles. The mildly singular nature of BPS hairy black holes,
studied in \cite{Markeviciute:2018yal,Markeviciute:2018cqs}, might be a clue for better
understanding their differences from the previous analytic solutions of
\cite{Gutowski:2004ez,Gutowski:2004yv,Chong:2005da,Kunduri:2006ek}. It may be
helpful to get a better notion on the near-horizon distinction of these two classes of
black holes. From the QFT dual side, it will be nice to develop a sharper criterion
for the hairiness of the deconfining saddle points. The condensations of certain
modes in the bulk force their dual operators to assume expectation values
at nonzero BPS chemical potentials. Within the simple consistent truncation of
\cite{Bhattacharyya:2010yg}, further studied in
\cite{Markeviciute:2016ivy,Markeviciute:2018yal,Markeviciute:2018cqs}, the
dual operator is easy to identify. With no guarantee that the deconfining saddle points
of this paper and of \cite{Choi:2018hmj} will be within this truncation ansatz, one should
figure out what kind of operators should be considered. Technically,
it is also interesting to see whether
one can find supersymmetric operators that can be inserted in the index.

It has been found in \cite{Choi:2018hmj} that the large charge limits of
non-hairy black holes \cite{Gutowski:2004ez,Gutowski:2004yv,Chong:2005da,Kunduri:2006ek}
are counted by the index. This presumably means that they are likely to be the dominant
saddle points in the large charge limit. It will be interesting to clarify how this
happens: for instance, whether there are further phase transitions to non-hairy black holes,
or whether hairy black holes asymptotically become indistinguishable with non-hairy ones.
For instance, we find some studies on large rotating AdS black holes
\cite{Dias:2010ma}, which can be made hairy only at very low Hawking temperature.
Although these are non-BPS black holes, they may give lessons to large BPS black holes.

\newpage


\hspace*{-0.8cm} {\bf\large Acknowledgements}
\vskip 0.2cm

\hspace*{-0.75cm} We thank Shiraz Minwalla for helpful discussions, especially
for comments which led to our section 2.3. SK thanks Ashoke Sen for discussions
in 2016, in which macroscopic boson/fermion cancelations were emphasized
to the author. We also thank Kimyeong Lee, Jaemo Park and Jorge Santos for helpful comments.
This work is supported in part by the National Research Foundation of Korea (NRF) Grant
2018R1A2B6004914 (SC, SK, JN), NRF-2017-Global Ph.D. Fellowship Program (SC),
and Hyundai Motor Chung Mong-Koo Foundation (JN).

\end{document}